\documentclass{aa}
\usepackage{txfonts}
\usepackage{natbib}
\usepackage{graphicx}
\bibpunct{(}{)}{;}{a}{}{,}

\begin{document}

\title{Identifying new opportunities for exoplanet characterisation at high spectral resolution}

\author{R.J.~de Kok\inst{\ref{int1}}   \and J.~Birkby\inst{\ref{int2}} \and M.~Brogi\inst{\ref{int2}} \and H.~Schwarz\inst{\ref{int2}} \and S.~Albrecht\inst{\ref{int3}} \and E.J.W.~de Mooij\inst{\ref{int4}} \and I.A.G.~Snellen\inst{\ref{int2}} }

\institute{SRON Netherlands Institute for Space Research, Sorbonnelaan 2, 3584 CA Utrecht, The Netherlands \email{R.J.de.Kok@sron.nl} \label{int1}  \and
Leiden Observatory, Leiden University, Postbus 9513, 2300 RA Leiden, The Netherlands \label{int2}  \and
Department of Physics, and Kavli Institute for Astrophysics and Space Research, Massachusetts Institute of Technology, Cambridge, Massachusetts 02139, USA \label{int3}  \and
Department of Astronomy and Astrophysics, University of Toronto, 50 St. George Street, Toronto, Ontario M5S 3H4, Canada \label{int4}  }

\date{}

\abstract
{Recently, there have been a series of detections of molecules in the atmospheres of extrasolar planets using high spectral resolution (R$\sim$100,000) observations, mostly using the CRyogenic high-resolution InfraRed Echelle Spectrograph (CRIRES) on the Very Large Telescope. These measurements are able to resolve molecular bands into individual absorption lines. Observing many lines simultaneously as their Doppler shift changes with time allows the detection of specific molecules in the atmosphere of the exoplanet.}{We aim to identify new ways of increasing the planet signal in these kinds of high-resolution observations. First of all, we wish to determine what wavelength settings can best be used to target certain molecules. Furthermore, we want to simulate exoplanet spectra of the day-side and night-side to see whether night-side observations are feasible at high spectral resolution.}{We performed simulations of high-resolution CRIRES observations of a planet's thermal emission and transit between 1-5 $\mu$m and performed a cross-correlation analysis on these results to assess how well the planet signal can be extracted. These simulations take into account telluric absorption, sky emission, realistic noise levels, and planet-to-star contrasts. We also simulated day-side and night-side spectra at high spectral resolution for  planets with and without a day-side temperature inversion, based on the cases of HD 189733b and HD 209458b.}{Several small wavelength regions in the L-band promise to yield cross-correlation signals from the thermal emission of hot Jupiters of H$_2$O, CH$_4$, CO$_2$, C$_2$H$_2$, and HCN that can exceed those of the current detections by up to a factor of 2-3 for the same integration time. For transit observations, the H-band is also attractive, with the H, K, and L-band giving cross-correlation signals of similar strength. High-resolution night-side spectra of hot Jupiters can give cross-correlation signals as high as the day-side, or even higher.}{We show that there are many new possibilities for high-resolution observations of exoplanet atmospheres that have expected planet signals at least as high as those already detected. Hence, high-resolution observations at well-chosen wavelengths and at different phases can improve our knowledge about hot Jupiter atmospheres significantly, already with currently available instrumentation.}


\titlerunning{Exoplanets at high spectral resolution}
\authorrunning{de Kok et al.}

\maketitle 

\section{Introduction}

The discovery of transiting extrasolar planets orbiting bright stars has opened up the possibility of characterising the atmospheres of these exoplanets through transmission and secondary eclipse spectroscopy. Broadband observations of transits and secondary eclipses have now become standard practice, both from space and the ground, with the transit and eclipse depths of many planets measured at a number of wavelengths \citep[e.g.][]{sea10}. On the other hand, real characterisation of exoplanet atmospheres is still in its infancy, since the data available for most planets contain only little information regarding the planet's gas abundances and temperature structure. Medium-resolution spectra can increase the available information on the planet significantly. However, some exciting results on the most favourable targets have subsequently been disputed \citep{swa10,man11, tin07, gib11,mad10,cro12}, highlighting the demanding nature of these kinds of observations \citep[see also][]{bea13}.

High-resolution spectroscopy, in which molecular bands are resolved into indivual absorption lines, can yield information on exoplanet atmospheres that is complementary to broadband transit and secondary eclipse measurements. Despite several attempts in the past \citep[e.g.][]{wie01,bar10}, high-resolution spectra have only recently provided their first detections of molecules on exoplanets. \citet{sne10} detected absorption by carbon monoxide lines during a transit of HD 209458b using the CRyogenic high-resolution InfraRed Echelle Spectrograph \citep[CRIRES,][]{kau04} on the Very Large Telescope, at a resolving power of R$\sim$100,000. A cross-correlation technique was used to obtain the planetary signal from the data by combining the signal of a few dozen CO lines. Interestingly, the signal of the CO lines was shifted 2$\pm$1 km s$^{-1}$ with respect to the system velocity, possibly indicating day-to-night winds. Subsequent modelling of the dynamics of HD 209458b indicates that such a Doppler shift in the planetary transmission spectrum might indeed be expected under certain conditions \citep{mil12,sho13}. Using CRIRES, absorption lines of CO have also been detected in the day-side thermal emission of the non-transiting planets $\tau$ Boo b \citep{bro12,rod12} and possibly 51 Peg b \citep{bro13}. In the latter case, water vapour was also possibly detected. These detections allowed the determination of the orbital inclination of these planets, and hence their true mass. Furthermore, it was found that these planets' temperature profiles do not show a strong inversion layer at the pressures probed by the CO lines (between $\sim$10$^{-5}$-1 bar), since the lines show up in absorption. The same conclusion was drawn for the transiting planet HD 189733b, using from CRIRES \citep{dek13} and Keck/HIRES \citep{rod13} observations. Subsequently, \citet{bir13} also found water vapour at 3.2 $\mu$m on this planet. There are large degeneracies in determining the temperature and molecular abundances from this type of high-resolution observations, mostly caused by the inability to determine the absolute continuum level, and by the general degeneracy between temperature profile and gas abundance. Nevertheless, constraints can be placed on the depth of the absorption lines, which should reduce uncertainties in the retrieved temperature and composition of transiting planets \citep{dek13}. Detections of multiple gases, and at multiple wavelengths, can also be used to determine relative abundances of gases, although the constraints set by \citet{dek13} and \citet{bir13} are not strict enough to constrain the gas abundances of CO and H$_2$O by themselves. During transit, the line depths can give more unambiguous information regarding the composition, especially when the continuum opacity source is known and well-defined, which constrains the absolute pressure levels that are probed (e.g. Rayleigh scattering at visible wavelengths or H$_2$-H$_2$ collision-induced absorption in the K-band. See also \citet{ben12}).

We aim to assess how high-resolution observations can best be used to improve our knowledge of exoplanet atmospheres, by seeing how best to detect specific molecules and how the planet signal can be measured at a higher signal-to-noise ratio in the near-infrared. We present simulations of high-resolution planet spectra and observations, in order to explore these new possibilities for  characterising exoplanet atmospheres. In Section 2, we investigate which wavelengths are best suited to search for a range of molecules, with a focus on the CRIRES instrument. In Section 3, we perform simulations to assess the detectability of the nightside atmosphere and day-night differences. In Section 4, we present conclusions.

\section{Optimal wavelengths}

Finding the best wavelengths to perform ground-based high-resolution observations of exoplanet atmospheres is far from trivial. On the one hand, a wavelength range should be chosen that has a strong exoplanet signal, caused by strong absorption lines. On the other hand, these strong absorption lines are often present in the Earth's atmosphere as well, greatly reducing the transmission of the Earth's atmosphere. Molecules that are not abundant in the Earth's atmosphere, but are abundant in the atmospheres of hot Jupiters, are therefore good candidates to observe, as illustrated by the recent detections of CO. We performed a sensitivity analysis on simulated data to quantify how well various molecular features can be observed, and from this determine what the best wavelengths to perform these kinds of observations are. We focused on the CRIRES instrument, since it has proven to be the most efficient for near-infrared high-resolution observations of exoplanets \citep[e.g.][]{dek13}, but lessons from this exercise can be applied to other instruments as well. Especially relevant will be the METIS instrument on the European Extremely Large Telescope \citep{bra12}. This instrument, which will greatly enhance the sensitivity of this method, will observe similarly fixed wavelength regions in the near-infrared. The main focus in this section is on the planet HD 189733b, since the results of the sensitivity analysis can be compared to our detections of CO and H$_2$O. Results for other planets are briefly discussed.

\subsection{Method}
The simulated data were created using ESO's Exposure Time Calculator\footnote{http://www.eso.org/observing/etc/bin/gen/... ...form?INS.NAME=CRIRES+INS.MODE=swspectr} (ETC), which simulates CRIRES' perfomance including a high-resolution model of the Earth's transmission and sky emission. The slit width was set to 0.2", which results in a resolution of R$\approx$100,000. The star was modelled as a blackbody, corresponding to the effective temperature of the planet's host star. This approach facilitated the insertion of a planet signal later in the simulations (see below). The target magnitude (K=5.54) was also taken from the known host star, which also served as the target star for the adaptive optics system. Typical values for the Paranal site for the seeing (1" at zenith) and precipitable water vapour column (2.5 mm) were used and a fixed airmass of 1.5 was assumed. A fixed integration time of 50 s was used for all wavelength settings to allow the comparison of the sensitivity for a fixed amount of observing time between the different setting. The ETC outputs the target spectrum ($y_t$), including telluric absorption, and the sky spectrum ($y_s$), which takes into account airglow lines and thermal background due to the telescope and atmosphere, separately. Parts of the spectra that are deemed unsuitable for scientific use, e.g.~in between spectral orders, are given zero values by the ETC. We took the target spectrum as the basis for our simulated observation. The noise ($\sigma$) was calculated as is normally done by the ETC, but without the small contributions from dark current and read-out noise:

\begin{equation}
\sigma=1.2 \sqrt{y_t + 2y_s}
\end{equation}

For the thermal emission spectra, we created a hundred spectra, each being the target spectrum with random Gaussian noise added, all with the above standard deviation, and inserted planet signals at phases between 0.3-0.7.  For transit observations, we created fifty spectra between phases -0.05-0.05. The planet spectra are the result of line-by-line model calculations, using HITEMP \citep{rot10} and HITRAN 2008 \citep{rot09} absorption data. A discussion about the use of line data is presented in Section 2.5. The calculated planet emission spectra were divided by the stellar blackbody spectrum used in the ETC, and scaled by $\left(\frac{R_p}{R_*}\right)^2$ to obtain the correct planet-to-star contrast. This planet signal was then multiplied by the ETC spectra and added to them to yield the simulated signal of the host star with an additional planet. The modelled transit spectra, which resulted from integrating over the entire planetary disc, are in units of fraction of starlight transmitted. We multiplied the ETC spectrum with these transit spectra to obtain the simulated transit observations.
We did not simulate the effect of changing airmass or throughput with time here to allow fair comparisons between different phases. In the end, we see very little variation in the correlation values for the different simulated spectra, and take the mean value in determining the efficiency of the planet detection. Since the only changing signal is now the Doppler shift of the planet, we can simplify our usual data analysis \citep{bro13,dek13}. `Instead of e.g.~fitting with airmass, we now simply subtracted the mean spectrum for each simulated observation. This should be equivalent to the real data with the stellar signals, telluric signals, and any additional broadband features removed. We cross-correlated the model planet spectrum, which are identical to the spectra that were inserted into the simulated spectra, with this reduced data at the known planet velocities. We did this both for data that have the planet signal inserted, and for the same data for which the planet signal was not inserted, to asses any potentially spurious cross-correlation signal created by the added noise. The difference in correlation between these two cases was then taken as a measure of the sensitivity of the measurement to the inserted planet signal. In practise, the cross-correlation signals from the added noise did not affect the relative planet signal for each wavelength setting significantly. Increasing the standard deviation of the noise did of course reduce the cross-correlation signal.
The data was cross-correlated with the same model spectrum that was inserted into the data, which implies we know the planet spectrum exactly and only concern ourselves with the efficiency of extracting this planet signal. This is of course not the case in real measurements, but in practise, performing the cross-correlation analysis with a wide range of model spectra, and selecting the best one, will reduce the differences between the real, disc-averaged, planet spectrum and the model spectrum.

\subsection{Results: Thermal emission}

\begin{figure*}[htp]
\centering
\includegraphics[width=17cm]{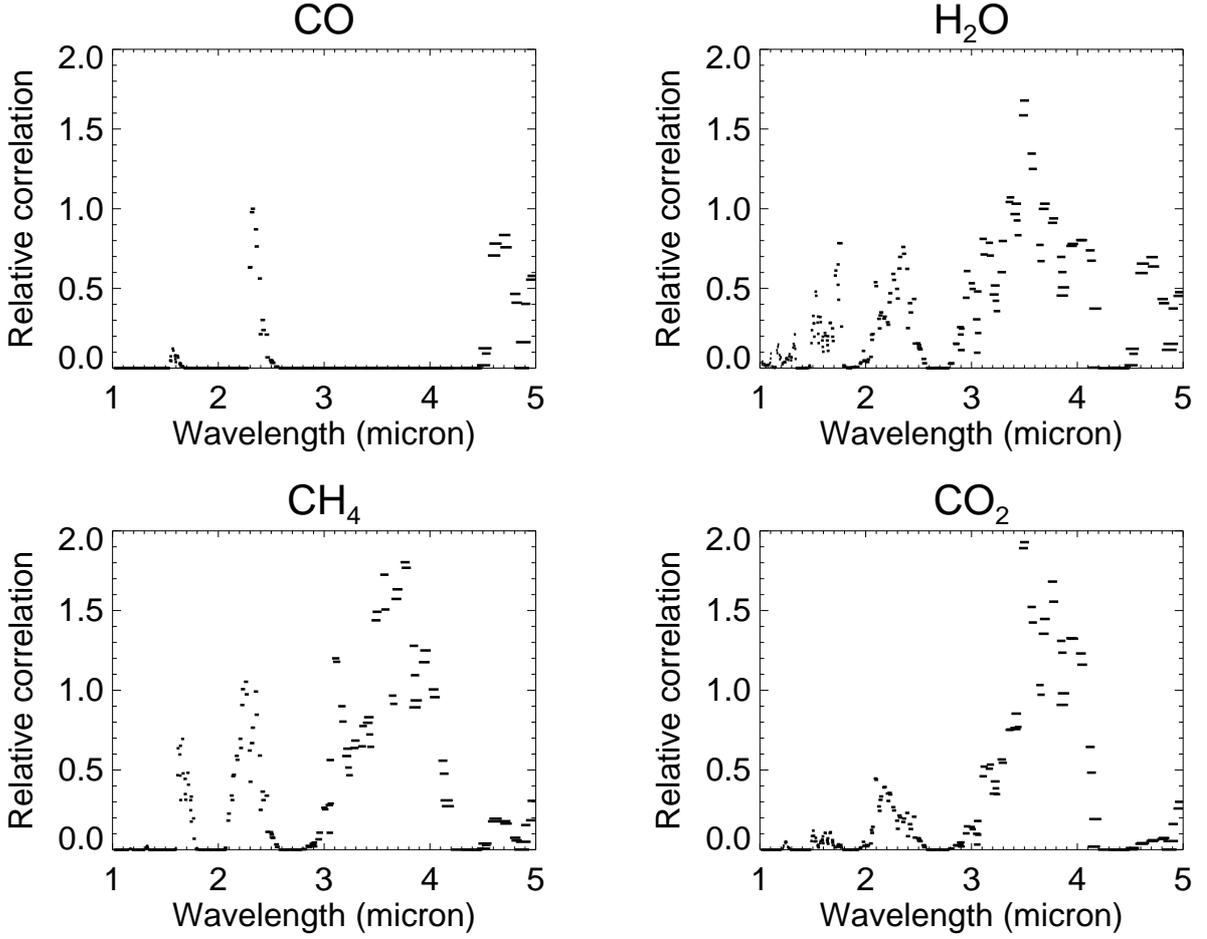}
\caption{Expected cross-correlation values, relative to the correlation of CO at 2.3 $\mu$m, for the different wavelength settings of CRIRES, for the thermal emission of a single gas in a hydrogen atmosphere.}
\label{fig.allmols}
\end{figure*}

The cross-correlations for thermal emission models with a single trace gas in a hydrogen-dominated atmosphere for the parameters of HD 189733b are shown in Fig.~\ref{fig.allmols}. The figure shows four spectroscopically active gases that are thought to be relatively abundant for HD 189733b from theory or measurements: H$_2$O, CO, CO$_2$ and CH$_4$. The temperature profile is chosen such that it is consistent with the retrieved profiles of \citet{mad09} and all molecules have an assumed volume mixing ratio of 10$^{-4}$. Simulations with more realistic abundances are discussed below. The cross-correlations are normalised by the cross-correlation of CO at 2.3~$\mu$m, which is where there is an already published detection \citep{dek13}. 

\begin{figure}[htp]
\centering
\resizebox{\hsize}{!}{\includegraphics{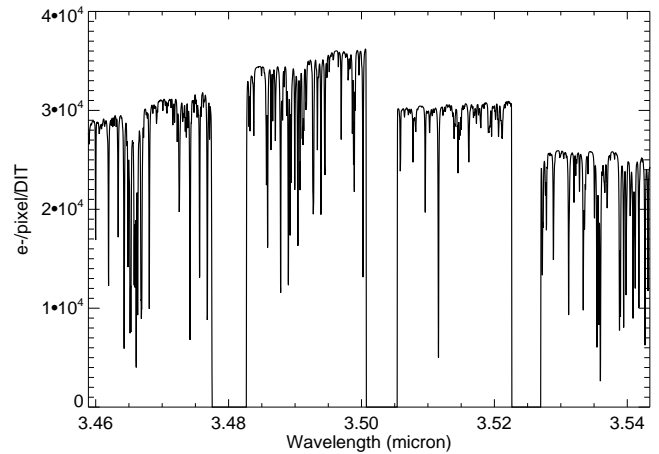}}
\caption{Output target spectrum from the ETC for the wavelength setting that gives the best expected signal for H$_2$O and CO$_2$. DIT is the integration time. All lines are telluric in nature.}
\label{fig.16}
\end{figure}

Fig.~\ref{fig.allmols} shows that CO is best observed with CRIRES around 2.3~$\mu$m for HD 189733b, where it was indeed observed. There is also an opportunity to observe CO in the M-band, where the CO lines are stronger, the planet-to-star contrast is higher, but the sky background becomes significant, and the flux from the planet is lower. Specific wavelength settings in the L-band generally provide the best opportunity to observe the thermal emission of the other gases, as plotted in Fig.~\ref{fig.allmols}. Especially around 3.5~$\mu$m the telluric transmission is very high and the balance between the flux from the planet and the noise from the star is optimal. At high temperatures, gases like CH$_4$, CO$_2$ and H$_2$O have many observable lines even where they are very weak in the Earth's atmosphere, meaning that the advantage of a high telluric transmission weighs more heavily than the reduced strengths of the absorption lines in the exoplanet's atmosphere. The target spectrum at 3.5 $\mu$m from the ETC, showing the telluric lines, are shown in Fig.~\ref{fig.16}. This shows that there are still telluric lines here, but these are generally not strong and leave much space for a potential planet signal. We note that the detection of H$_2$O by \citet{bir13} was achieved at 3.2 $\mu$m, where we predict the H$_2$O signal-to-noise to be rougly a factor of 2-3 smaller than at 3.5 $\mu$m. In the H-band, the planet flux is reduced and the expected signal is generally lower than in the K-band.

\begin{figure*}[htp]
\centering
\includegraphics[width=17cm]{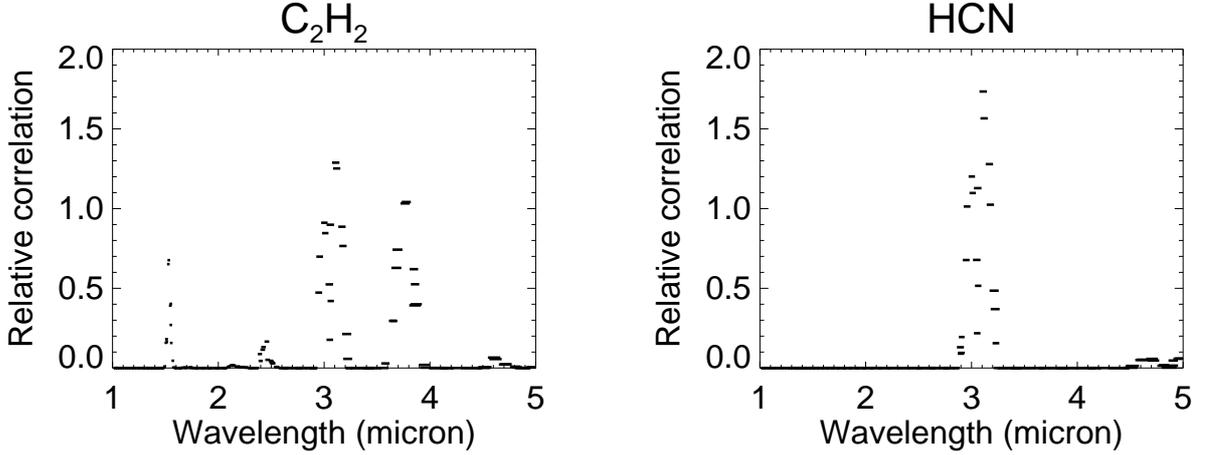}
\caption{As Fig.~\ref{fig.allmols}, but now for C$_2$H$_2$ and HCN.}
\label{fig.allmolsa}
\end{figure*}

We performed the same analysis for the molecules C$_2$H$_2$ and HCN. These gases are predicted to be abundant if photochemistry is important, or if the planet has a high C/O ratio \citep{mos13}. We note that the line data for these gases are very incomplete in HITRAN, with most notably many weak lines missing, which can become more important at higher temperatures. Self-consistent, more complete linelists for these gases will be available in the near-future (S.~Yurchenko, pers.~comm.), but we limit ourselves to the HITRAN list for now. Fig.~\ref{fig.allmolsa} shows another promising wavelength region around 3.1 $\mu$m where the carbon-bearing molecules CH$_4$, C$_2$H$_2$, and HCN all show a relatively good signal.

\begin{figure}
\centering
\resizebox{\hsize}{!}{\includegraphics{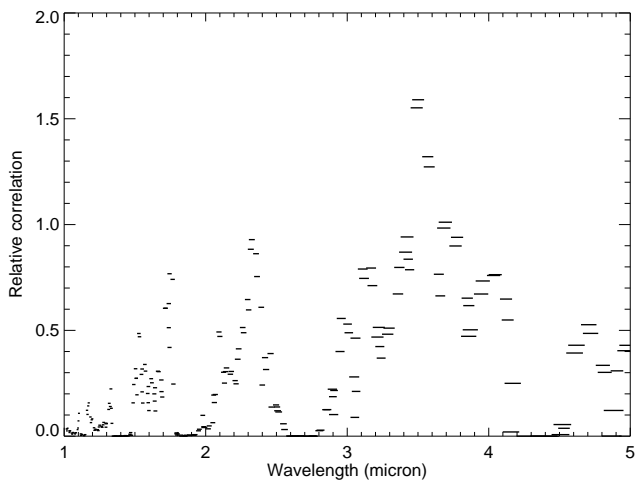}}
\caption{As Fig.~\ref{fig.allmols}, but now for a model spectrum with H$_2$O, CO, CH$_4$, and CO$_2$ (see text).}
\label{fig.allmolsall}
\end{figure}

In a real atmosphere there will be a combination of signatures from different molecules. The expected signal from spectra of HD 189733b with more realistic gas abundances is shown in Fig.~\ref{fig.allmolsall}. The volume mixing ratios of the gases were guided by \citet{mos13} (green lines in their Fig.~2): [H$_2$O] = 10$^{-4}$, [CO] = 10$^{-4}$, [CO$_2$] = 10$^{-6}$, and [CH$_4$] = 2$\times$10$^{-6}$. The best obtainable signal, assuming a perfectly accurate planet spectrum, again lies at 3.5 $\mu$m, and is actually very similar to the results for pure water. This is not surprising, since the spectrum is dominated by water lines in this case. However, when performing the cross-correlation analysis with a spectrum from only one gas, which is then different from the spectrum that is inserted into the data in this case, we see that the strong CH$_4$ lines can have a shielding effect at 3.5 $\mu$m, reducing the signal by up to a factor of two for high CH$_4$ volume mixing ratios. H$_2$O has a slightly less strong shielding effect. Of course, including CH$_4$ in the template spectrum increases the correlation signal again. Decreasing the CH$_4$ volume mixing ratio increases the ease with which the other molecules are detected in that wavelength region. So, although H$_2$O and CH$_4$ can have a shielding effect, it does not prevent the detection of other molecules as long as there are lines from the other molecule that are stronger than the H$_2$O or CH$_4$ absorption at that wavelength. For the assumed temperature profile, the expected cross-correlation signal decreases by a factor of 2-3 for every order of magnitude the volume mixing ratio of H$_2$O and CO$_2$ is decreased from 10$^{-4}$ downwards to 10$^{-7}$. 

In all the above analysis, the observing conditions, such as seeing and water vapour content, were not ideal. We repeated some of the calculations for more ideal situations, but this did not qualitatively change the relative strengths of the signals for the different wavelength settings. Quantitatively, the cross-correlation signal increased by $\sim$20\% in the ideal case. This is not surprising, since the best wavelength settings are already those least affected by the Earth's atmosphere.

\begin{figure}[htp]
\centering
\resizebox{\hsize}{!}{\includegraphics{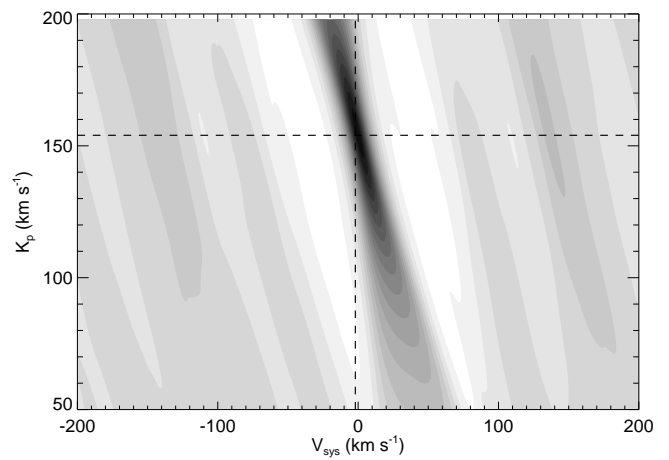}}
\caption{Cross-correlation values as a function of $K_p$ and $V_\mathrm{sys}$ for a simulated observation run at 2.3 $\mu$m, targeting CO. Normalised values range from -0.2 to 1. Dashed lines indicate the position of the inserted planet signal.}
\label{fig.acor}
\end{figure}

In our previous detections of the thermal emission of hot Jupiters \citep{bro12,bro13,dek13,bir13}, we showed the correlation values for a grid of planetary radial velocity amplitudes ($K_p$) and systemic velocities ($V_\mathrm{sys}$), to show that the detected signal is not an outlier and lies at the expected position in this grid. The significance of the detection was also partly evaluated using this grid. For non-transiting planets, the correlation peak in this grid was used to determine the inclination of the system. It is therefore interesting to see what the correlation values for such a grid are in our simulations. Again, we calculated the difference in correlation values for data with and without an inserted planet signal, but now for a smaller phase range (0.38-0.48), which is closer to the real measurements during a single night, for the grid of $K_p$ and $V_\mathrm{sys}$. The results are shown in Fig.~\ref{fig.acor}, for CO at 2.3 $\mu$m. Very similar patterns are seen here as are seen for the real data. Most notably, the maximum signal is elongated along slanted lines in the grid. This arises because the planet radial velocities during a small part of the orbit can be well reproduced by a range of $K_p$ and $V_\mathrm{sys}$ combinations. When two nights are used that probe both sides of phase 0.5, a cross-hatch pattern emerges \citep[see][]{bro13}. Fig.~\ref{fig.acor} also shows negative wings around the peak correlation and positive peaks farther away in $V_\mathrm{sys}$. These are the result of the correlation of the model spectrum with itself at a range of Doppler shift (i.e.~the autocorrelation function). In our previous work, these structures are also included in our determination of the correlation noise, which determined the significance of the detection, whereas in fact they can be seen as a reflection of the planet signal. 

\subsection{Results: Transmission}

\begin{figure*}
\centering
\includegraphics[width=17cm]{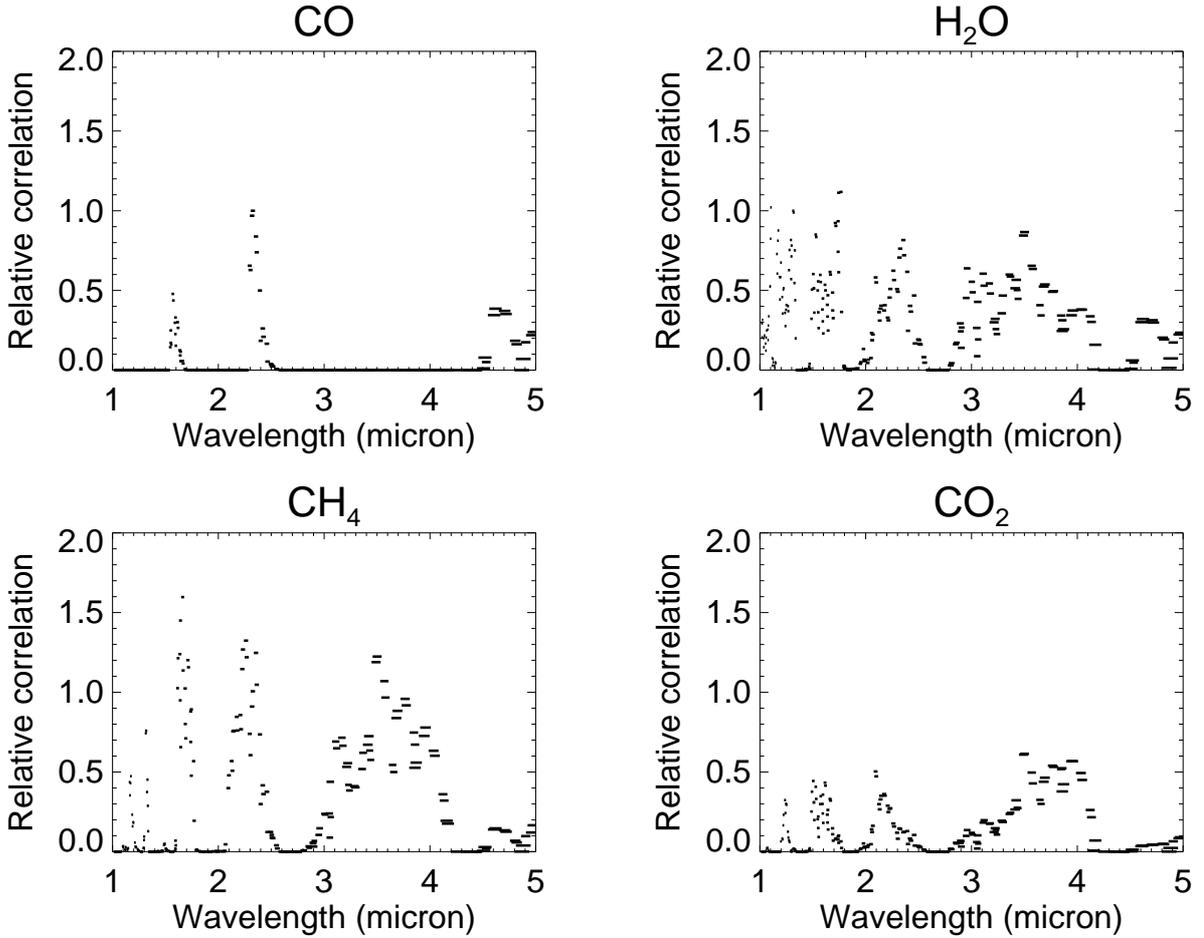}
\caption{Expected cross-correlation values, divided by the correlation of CO at 2.3 $\mu$m, for the different wavelength settings of CRIRES, for the transit spectrum of a single trace gas in a hydrogen-dominated atmosphere.}
\label{fig.allmolst}
\end{figure*}

\begin{figure*}
\centering
\includegraphics[width=17cm]{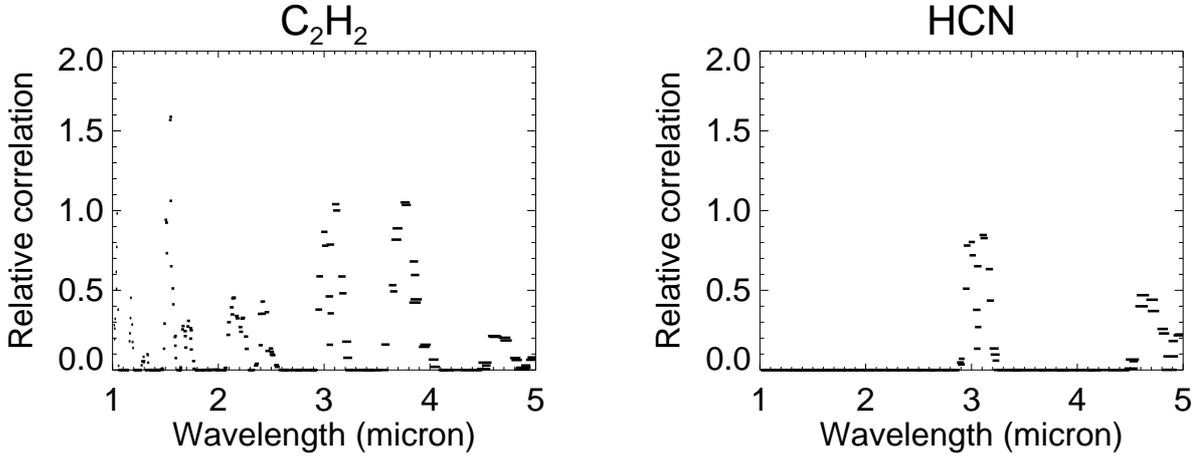}
\caption{As Fig.~\ref{fig.allmolst}, but now for C$_2$H$_2$ and HCN.}
\label{fig.allmolsta}
\end{figure*}

The analysis for transit signals of HD 189733b is shown in Figs.~\ref{fig.allmolst} and \ref{fig.allmolsta}. Here, shorter wavelengths are more favourable compared to the thermal emission spectra, since the planet signal is to first order dependent on the atmospheric scale height, which does not depend on wavelength. Hence, it does not drop quickly towards shorter wavelengths, unlike the Planck function at the exoplanet temperatures. The expected signals are therefore more similar between the different atmospheric windows. The comparative signal of a single molecule at multiple wavelengths can be used to search for haze and cloud signals, since the molecular lines will be less deep when haze is present.

As can be seen in Fig.~\ref{fig.allmolsta}, a transit measurement around 1.5 $\mu$m might be an excellent way to probe C$_2$H$_2$, which is predicted to be very abundant in the upper atmosphere for planets with a high C/O ratio. We note that this wavelength is also covered by HST/WFC3. Other molecules also give good signals in the H-band for these transit simulations. A decrease in the volume mixing ratio of a factor of 10 gives a decrease in cross-correlation signal of a factor of $\sim$2.

\subsection{Other planets}

In the above simulations we focused on HD 189733b. For other hot Jupiters, the best wavelength setting to observe a certain molecule will generally be the same, given a clear atmosphere and a lack of temperature inversion. There will be a difference in the expected signal when two wavelengths are considered that are far apart. For instance, the 3.5 $\mu$m H$_2$O signal  for $\tau$ Boo b and 51 Peg b is not as strong relative to the 2.3 $\mu$m CO signal compared to HD 189733b. However, the 3.5 $\mu$m H$_2$O signal is still higher than the CO signal if the water abundance is as high as the CO abundance. These differences are present, because, generally, the relative correlation signal as a function of wavelength from one planet to the next roughly scales with the emission or transmission signal from the planet, divided by the square root of the emission of the star, since the star is the source of photon noise. Differences in the temperature structure, gas abundances and haze opacity will give rise to further differences in the expected correlation signals.

\subsection{Practical considerations}

\begin{figure}[htp]
\centering
\resizebox{\hsize}{!}{\includegraphics{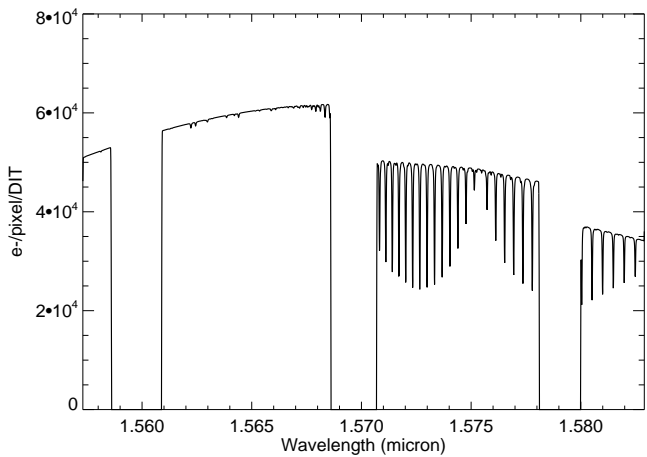}}
\caption{Output target spectrum from the ETC for the wavelength setting that gives a the best expected signal for CO in H-band. DIT is the integration time.}
\label{fig.36}
\end{figure}

Besides the theoretical ease of detecting a certain molecule with high-resolution spectroscopy, as discussed above, there are other aspects that can hinder detection. For instance, an accurate wavelength solution for the measured spectra is crucial in obtaining a good signal from the cross-correlation technique. In \citet{sne10,bro12,bro13,dek13}, we used telluric absorption lines to this end. Figs.~\ref{fig.16} and \ref{fig.36} show two spectra from the ETC in regions of high transmission in the L-band and the H-band, which are very suitable for observations of H$_2$O and CO respectively. The figures show that in the H-band telluric absorption lines might be very sparse at some wavelength regions, and one might need to rely on stellar lines, sky lines, or additional wavelength-calibration measurements for an accurate wavelength solution throughout the duration of the observation. The latter would result in a significant reduction in efficiency of the measurements, since the wavelength calibration will need to be performed several times during the course of a night. We note that the CO$_2$ lines in the right part of Fig.~\ref{fig.36} are actually very suitable for the wavelength-calibration there. The telluric spectrum from Fig.~\ref{fig.16} shows a nice spread of lines across all wavelengths, which would be very suitable for the wavelength-calibrations in this wavelength setting. Changes in the telluric lines, due to, for instance, changes in the water vapour content or seeing, are generally well removed in the data reduction process before cross-correlating with the model spectra.

The cross-correlation method for these high-resolution observations is also very sensitive to our exact knowledge of the line positions of the many lines in the exoplanet spectrum. If there is a random error in these line positions, the cross-correlation signal will be smeared out and sensitivity will be lower. This is an issue when using ab initio calculations of molecular lines, which are often necessary to calculate spectra for high temperatures, in particular for lines that are weak at lower temperatures. Especially the line intensities extrapolated to high temperatures from Earth-focussed databases, like HITRAN, can have large errors. On the other hand, ab initio calculations can have larger errors in the line positions \citep[see e.g.][]{bai12}. For high-resolution observations of certain less-studied molecules, the choice of absorption data can thus be a choice  between completeness and accurate line intensities on the one hand, versus accurate line positions on the other. In the (near) future, initiatives like Exomol \citep{ten12} and HITEMP \citep{rot10} will improve on this situation. In this paper, we used HITRAN for molecules that were not in HITEMP, with the aforementioned disadvantages noted.

Another issue that has been ignored in the previous calculations is the presence of stellar absorption lines. These show small (sub-pixel) Doppler shifts due to gravitational pull of the planet. This will result in relatively large residuals in our usual data reduction, which aims at removing signals that do not show any changes in barycentric velocity with time. When the targeted molecule in the planet atmosphere is also present in the stellar atmosphere, these residuals in the data can give rise to relatively strong cross-correlation signals, which hinder the detection of the planet signal near phases 0.0 and 0.5 \citep[e.g.][]{bro13}. Hence, during transit observations the stellar lines will be especially bothersome and more steps in the data reduction will need to be taken to reduce the effect of stellar lines, such as modelling and removing the stellar lines (Brogi et al., in prep.). 

\section{Night-side observations}

One commonality among the present detections of the thermal emission of exoplanet atmospheres at high resolution is that they probe the day-side, close to phase 0.5. Detecting the high-resolution planet signal throughout its orbit could reveal differences in the composition and temperature structure between the day-side and the night-side. Spitzer observations \citep[e.g.][]{knu09,knu12} reveal that the night-side of a hot Jupiter is generally colder and emits less broadband flux. Models of the dynamics of hot Jupiter atmospheres \citep[e.g.][]{sho09,dob13,rau13} indicate that at low latitudes the day-night differences are largest in the upper atmosphere, at the millibar level, where radiative time scales are short, and where the cores of absorption lines probe in high-resolution observations. Deeper in the atmosphere, where the continuum of high-resolution spectra probe, the day-night differences are generally smaller. This implies that in high-resolution spectra the depth of the lines at the night-side are not necessarily less deep than at the day-side, since the day-night temperature differences probed by the line cores are larger than those probed by the continuum. 

\begin{figure}[htp]
\centering
\resizebox{\hsize}{!}{\includegraphics{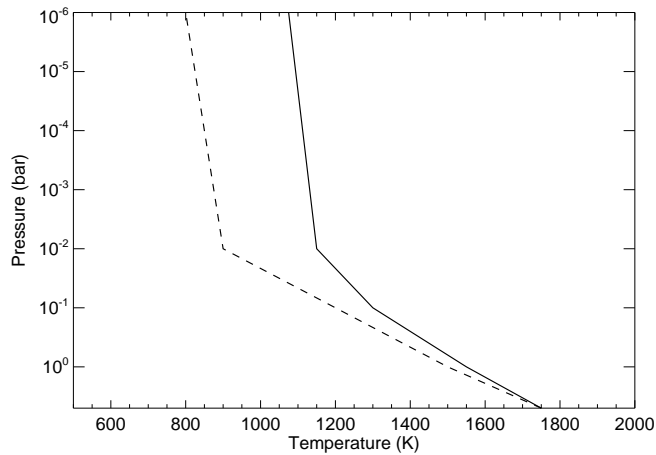}}
\caption{Assumed temperature profiles for the day-side (solid line) and night-side (dashed line) of HD 189733b.}
\label{fig.daynighttemphd189}
\end{figure}

\begin{figure}[htp]
\centering
\resizebox{\hsize}{!}{\includegraphics{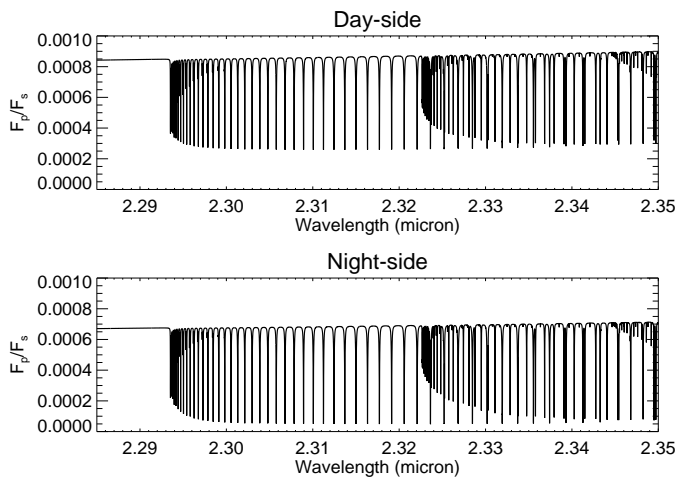}}
\caption{CO spectra, in units of planet-star contrast, for approximated day-side and night-side temperatures of HD 189733b (see Fig.~\ref{fig.daynighttemphd189}). }
\label{fig.daynighthd189}
\end{figure}

To explore the differences in high-resolution spectra between the day-side and the night-side we calculated spectra with temperature structures based on dynamical model calculations from the literature. Fig.~\ref{fig.daynighttemphd189} shows day-side and night-side temperatures of HD 189733b based on low latitude profiles from \citet{sho09}. The corresponding spectra, for an atmosphere with only H$_2$, and CO at volume mixing ratio of 10$^{-4}$, are plotted in Fig.~\ref{fig.daynighthd189}. The two spectra are very similar, with the main difference being an offset between the spectra. Despite the lower overall flux level of the night-side spectra, the depths of the lines are actually almost identical in this case. The lines in the night-side in fact represent a larger difference in brightness temperature between the continuum and the line cores, since at lower temperatures a larger brightness temperature shift is needed to obtain the same shift in flux. 

\begin{figure}[htp]
\centering
\resizebox{\hsize}{!}{\includegraphics{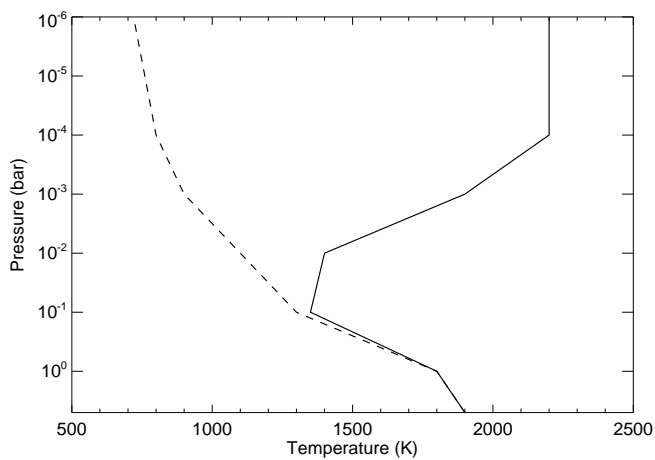}}
\caption{Assumed temperature profiles for the day-side (solid line) and night-side (dashed line) of HD 209458b.}
\label{fig.hd209temp}
\end{figure}

Another example explored is HD209458b. This planet is thought to have a strong temperature inversion at the dayside, which presents itself as emission lines in the spectrum.  To get an impression of how such a planet with a temperature inversion would manifest itself in correlation analyses across its orbit, we calculated spectra at a range of emission angles to create a simple spatially inhomogeneous planet for which disc-integrated spectra were calculated along its orbit. The three-dimensional structure of the planet was strongly approximated, based on \citet{par13}, by assuming a day-side temperature profile along a 140$^\circ$$\times$140$^\circ$ area centred around the sub-solar point and a night-side temperature profile everywhere else. The temperature profiles for these two regions are plotted in Fig.~\ref{fig.hd209temp}. At low and high pressures, such an approach seems reasonable, given the dynamical model outputs for HD 209458b \citep[e.g.][]{sho09,par13}. At intermediate pressures the real planet atmosphere is more complicated than assumed here, but it is sufficient to obtain a qualitative picture of the possible high-resolution spectra along the planet's orbit.

\begin{figure}[htp]
\centering
\resizebox{\hsize}{!}{\includegraphics{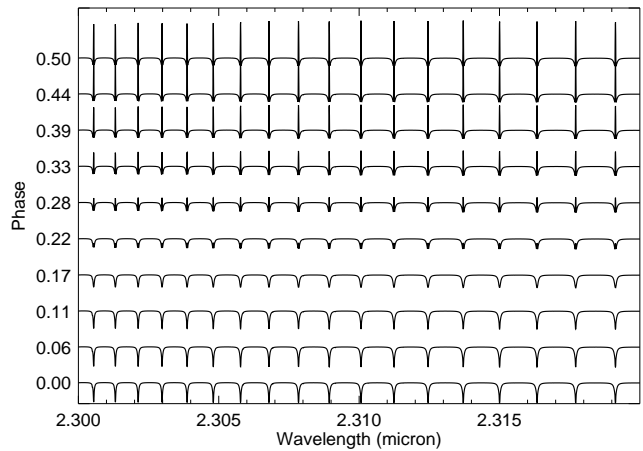}}
\caption{Simulated CO spectra for HD 209458b across half its orbit for a day-side region with a temperature inversion.}
\label{fig.hd209}
\end{figure}

The disc-integrated spectra for phases between 0 and 0.5 are shown in Fig.~\ref{fig.hd209}. As can be seen, at intermediate phases the contibutions from the inverted and non-inverted parts of the atmosphere can cancel out for a large part, greatly reducing any potential signal at high resolution. In all cases, the wings of CO lines show up in absorption, since they probe the part of the atmosphere that is still getting cooler with increasing altitude. Depending on the exact location of the temperature minimum in the inverted temperature profile, these wings will be a strong or weak feature of the disc-integrated spectrum between phases $\sim$0.25-0.75. We note that high-resolution observations will generally be more suitable to determine whether the temperature profile has an invertion than broadband measurements, since the line shapes are probed directly.

For an irradiation level that is in between HD 189733b and HD 209458b, the day-side atmosphere might be close to isothermal at pressures probed by molecular absorption lines. The secondary eclipse depths of several exoplanets also resemble those of blackbodies, which could indicate a day-side atmosphere that is close to isothermal (e.g. WASP-18b \citep{nym11}, WASP-12b \citep{cro12}, TRES-3b \citep{cro10}, HAT-P-1b \citep{tod10}, WASP-24b \citep{smi12}, etc.). For such planets, the night-side might yield much deeper absorption lines and hence a much larger high-resolution signal, especially if the horizontal heat distribution at low pressures is known to be inefficient \citep{cow12,max13}. We also point out that a single night of CRIRES measurements of $\tau$ Boo b yields a weaker CO signal \citep{bro12,rod12} than HD 189733b \citep{dek13}, despite the planet being hotter, and orbiting a nearer star. Hence, $\tau$ Boo b might also be a planet with a relatively small temperature gradient at the day-side and night-side observations can possibly yield a larger correlation signal.

\section{Summary and conclusions}

We simulated exoplanet spectra at high spectral resolution to assess how high-resolution observations can be used to further improve our knowledge of exoplanet atmospheres. We applied a cross-correlation technique on simulated spectra from the CRyogenic high-resolution InfraRed Echelle Spectrograph (CRIRES) on the Very Large Telescope from which we identified specific spectral regions that give the highest signal when measuring hot Jupiter spectra at high spectral resolution. For detections of the thermal emission, specific regions in the L-band give the best results for H$_2$O, CO$_2$ and CH$_4$, as well as C$_2$H$_2$ and HCN. In this respect, the planned infrared instrument on the European Extremely Large Telescope, METIS \citep{bra12}, which includes a high-resolution channel in the L-band, is expected to give excellent signals for hot Jupiters and will also be able to study colder objects in this way. For high-resolution transmission signatures shorter wavelengths also give good signals, giving the opportunity to detect the same molecule at multiple wavelengths. This would give constraints on haze opacity and possibly vertical variations in temperature and gas abundances.

Other future instruments that show promise in detecting molecules in exoplanet atmospheres at high resolution in the near-infrared are CARMENES \citep{qui12} and SPIRou \citep{thi12}. These spectrographs have a lower spectral resolution (R=75,000 for SPIRou and R=82,000 for CARMENES), and are planned on smaller telescopes ($\sim$3.5 m diameter). However, they have very accurate wavelength solutions, necessary to perform stellar radial velocity determinations, and a higher throughput. This will benefit the detection method discussed in this paper. They also have a much wider instantaneous wavelength coverage, with CARMENES covering the YJH-bands, and SPIRou the YJHK-bands. Although the lower spectral resolution will decrease the line depths of the planet spectrum and the ability to separate the planet signal from telluric lines, the increase in wavelength coverage implies that many more exoplanet absorption lines can be observed simultaneously, increasing the correlation signal. Whether CARMENES and SPIRou can detect molecules in exoplanet atmospheres more or less efficiently than CRIRES is not immediately clear and will need further investigation. In this light, the increase of wavelength coverage obtained by turning CRIRES into a cross-dispersed spectograph \citep{oli12} is very exciting and can improve the efficiency of exoplanet atmosphere detections with CRIRES by a factor of several. 

We also calculated example spectra of the day-side and night-side of the planets HD 189733b and HD 209458b at high resolution. These indicate that, despite a lower broadband flux, the night-side does not necessarily give rise to shallower absorption lines at high spectral resolution. For planets with a weak day-side temperature gradient, night-side observations may actually be more suitable for high-resolution observations than the day-side. High-resolution observations covering the entire orbit of the planet can give great insight into the three-dimensional temperature structure and chemistry of hot Jupiters and greatly complements the information from Spitzer phase curves.

\begin{acknowledgements}
We thank the anonymous referee for his/her useful comments and suggestions. This work was funded by the Netherlands Organisation for Scientific Research (NWO).

\end{acknowledgements}

\end{document}